\documentclass[twocolumn,amssymb,amsmath,floats,showpacs,pre]{revtex4-1}
\usepackage{amsmath,amssymb,bm}
\usepackage{graphicx}
\usepackage{xcolor}
\usepackage[toc,page]{appendix}
\usepackage{comment}
\usepackage[unicode=true]{hyperref}
\usepackage{enumerate}
\graphicspath{{images/}}

\begin{document}

\title{Control of cascading failures using protective measures}

\author{Davood Fazli$^{1}$, Mozhgan Khanjanianpak$^{2}$}
\author{Nahid Azimi-Tafreshi$^{1}$}
\email{nahid.azimi@iasbs.ac.ir}
\affiliation{$^1$ Physics Department, Institute for Advanced Studies in Basic Sciences (IASBS), Zanjan 45137-66736, Iran \\ $^2$ Pasargad Institute for Advanced Innovative Solutions (PIAIS), Tehran 1991633357, Iran}

\date{\today}

\begin{abstract}
Cascading failures, triggered by a local perturbation, can be catastrophic and cause irreparable damages in a wide area. Hence, blocking the devastating cascades is an important issue in real world networks. One of the ways to control the cascade is to use protective me‌asures, so that the agents decide to be protected against failure. Here, we consider a coevolution of the linear threshold mo‌del for the spread of cascading failures and a decision-making game based on the perceived risk of failure. Protected agents are less vulnerable to failure and in return the size of the cascade affects the agent's decision to get insured. We find at what range of protection efficiency and cost of failure, the global cascades stop. Also we observe that in some range of protection efficiency, a bistable region emerges for the size of cascade and the prevalence of protected agents. Moreover, we show how savings or the ability of agents to repair can prevent cascades from occurring.
\end{abstract}

\maketitle

\section*{Introduction}

Cascade dynamics occurs in a wide range of real-world networks. The spread of rumors, failures, cultural fads, innovation and so on, may occur as a cascade or avalanche in social, economics and technological networks \cite{fasion, rumor, information, social22, failure}. The cascading effects start from a local perturbation and due to the interactions between agents of a network, propagate through the whole of the network as a domino-like process.

The linear threshold model, proposed by Watts, is a well-known model in the field of complex contagion which describes the cascade dynamics \cite{Watts}. The process is initially started from a configuration in which a few randomly chosen nodes are set to be active. The activation of interacting inactive nodes is determined by the state of the active neighbor nodes according to a simple threshold rule: At each time step, each inactive node updates its state active if the fraction of its active neighbors exceeds a threshold value, otherwise it remains inactive. Once being active, a node cannot turn back to the inactive state. The dynamics continues until no node changes its state. The threshold model has been extensively used to model information cascades on social networks \cite{social, social1, social2, social3}, contagion in financial networks \cite{financial, financial2} and cascading failures in different various networks, e.g., multilayer networks \cite{tec1, tec2} and hypergraphs \cite{hypergraph}. It was shown that the size of cascade depends on the size of the initially active nodes (seed size) \cite{seedsize, initial}, the selection strategy of the initiators \cite{social3} and distribution of threshold values among nodes \cite{heterogenous}. Also the structure of the underlying network can affect the dynamics of cascade \cite{modular, clustered, weighted, multiplex1, multiplex2, multiplex3}.

Cascade propagation can sometimes be destructive and cause catastrophic effects throughout the network. For instance, in economic networks the insolvencies of some financial institutions can spread sequentially to other institutions with inadequate capital levels and create a global financial crisis \cite{financial3, financial4}. Also, in an electrical power grid, an initial failure or power outage in one area can spread to the surrounding area and eventually affect the entire network that could lead to a catastrophic collapse of the system \cite{power}. Traffic congestion in transportation systems \cite{trans1, trans2} and mass extinction in ecosystems \cite{ecology} are other examples of catastrophic cascades. Hence developing strategies to prevent and recover the system from such failures is a challenging problem.

One of the strategies to halt and reduce the cascade propagation is to block or protect some of the nodes, such that they resist adopting the active state. For instance, in spreading of innovations some individuals may refuse to adopt because of interest in another product or the effect of the external factors such as the media. Insurance or financial support of the government can be also effective in spreading the bankruptcy of companies. To study the effect of blocked nodes on cascade formation, Ruan et. al., extended the threshold model by considering that some nodes are blocked, so that they never adapt, and showed that with increasing fraction of blocked or immune nodes, the speed of spreading slows down and the global cascade size reduces \cite{Ruan}. Identification of influential nodes, which are able to form activities in larger populations, and protect‌ing them can be more effective to control the entire network’s dynamics \cite{influential1, influential2}.

However the protection is costly. Hence, the individuals or companies must decide to pay a cost to protect and consequently reduce the spread of cascade or accept the risk of failure. The decision making process can be studied in the framework of the theoretical game theory, such that each agent tends to maximize his own payoff according to the cost-benefit ratio associated with protective measures \cite{game, game2, game3, game4}. The agents' decision to protect affects on the spreading of cascade and on the other hand decision making depends strongly on the cascade dynamics and the perceived risk of failure. Hence we need to consider a coevolution of two dynamics.

In this paper, we couple the threshold model for the spread of cascade and a decision game based on the perceived risk of failure. Using this framework, we study the effect of protected nodes in controlling cascading failures. In this way, using numerical simulations, we find the phase diagram of the model and show if the protection efficiency and also the cost of joining the cascade are smaller than a certain limit, there is no desire to protect and the avalanche of failures covers a significant part of the network. We observe that there is a bistable region for certain values of parameters in the phase diagram of the model.

In real-world systems, damaged components can sometimes repair themselves. For instance, the individuals or financial companies have sometimes the ability or financial reserves to improve the situation and return to the initial situation in case of failure. Also an intersection in a road network may be overloaded and in congestion in peak hours but the intersection traffic can be restored by reducing the traffic pressure. In the case of recovery possibility, the formation of global cascades can be avoided. Here we also consider the possibility of recovery in our model and find that above a critical value for the recovery probability, the cascade stops.

The article is organized as follows. In the next section, we define the dynamics of the model on a random network. In Sec.~\ref{sec3} we discuss the interplay between the game dynamics and the threshold model dynamics and find the role of protection efficiency, cost of failures and the recovery probability in the control of cascades. The paper is concluded in section \ref{secCon}.

\section*{\label{sec2}Materials and methods }
\subsection*{Model}

We consider a society of size $N$, such that the agents are interacting with each other through a random network. An activation process propagates on the network following the dynamics of the linear threshold model, in which the nodes are active (A) or inactive (I). Let us denote the state of node $i$ at time $t$ with $s_i(t) \in \{ 0,1\}$, such that $s_i(t)=0$ if the node is inactive and $s_i(t)=1$ if it is active.
Initially there are a few fraction of active nodes as seed nodes and others are inactive.
At each time step $t$, the state of nodes are updated synchronously. Thus, an inactive agent becomes active if the fraction of its active neighbors is at least $m$. In the standard threshold model it is assumed that whenever a node becomes active, it remains active forever. However we assume a reversible threshold model, such that an active node will be again inactive with probability $\beta$. This modification allows the active (damaged) nodes to recover. Updating process continues until the dynamics reaches a stationary state. The fraction of active nodes at the stationary state is considered as the size of cascade.
\begin{figure}[t!]
	\includegraphics[scale=0.3]{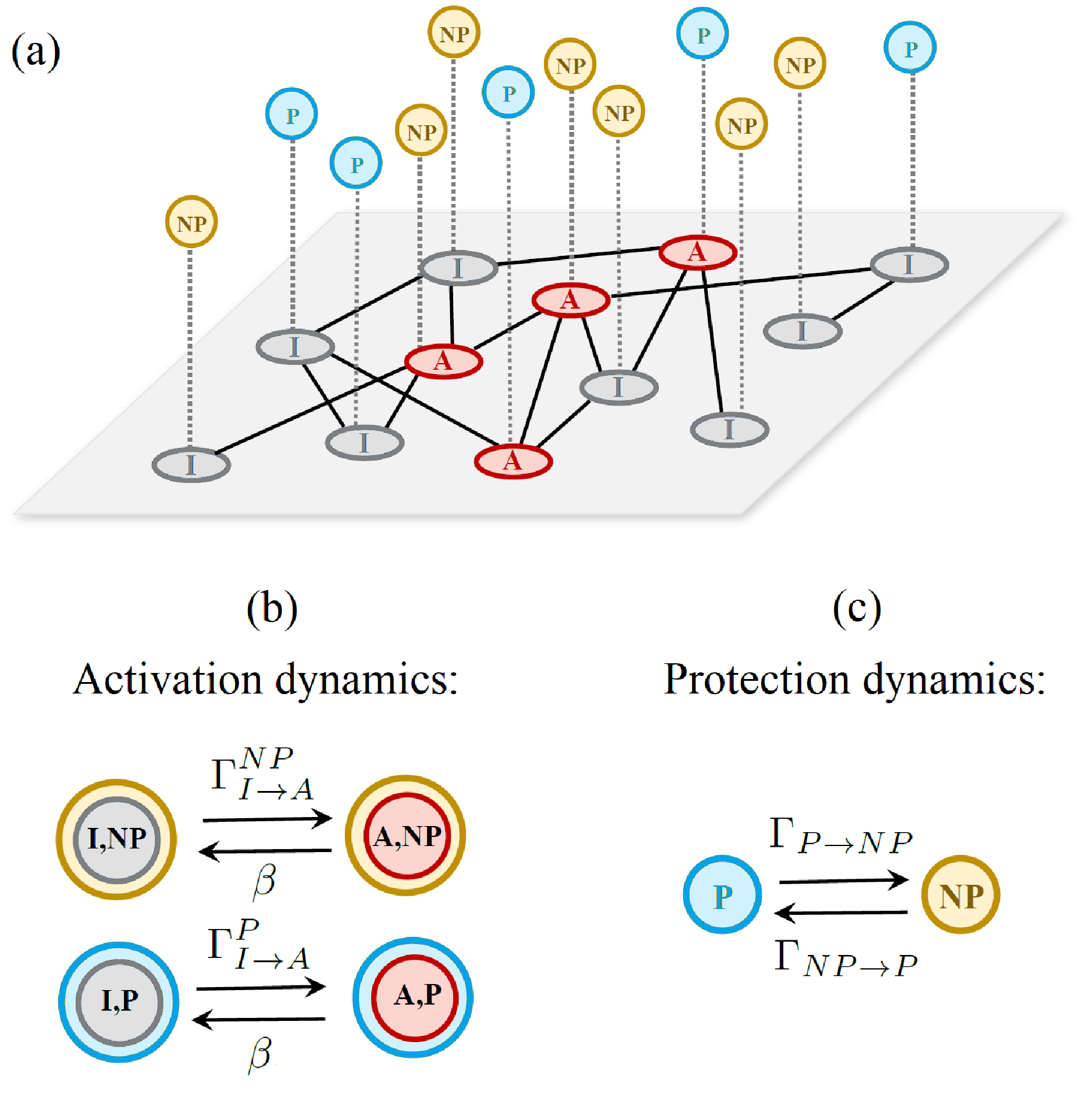}
	\centering
	\caption{Schematic of the model with transition probabilities of coupled dynamics. (a) The network of connections where gray and red nodes indicate inactive (I) and active (A) nodes, respectively. Each node may be protected (P) or not-protected (NP), shown by blue and cream circles. (b) An inactive and protected (I,P) or inactive and not-protected (I, NP) node becomes active according to the threshold model with probabilities $\Gamma^{P}_{I\to A}$ and $\Gamma^{NP}_{I\to A}$, respectively. Active nodes return to inactive state with a constant probability $\beta$. (c) Nodes choose the strategy protected or not-protected based on a decision game with probabilities $\Gamma_{NP\to P}$ and $\Gamma_{P\to NP}$, respectively. }
	\label{model}
\end{figure}

To overcome the global cascades, we consider the possibility that some agents decide to insure themselves against breakdowns. Insured agents with paying a cost are less exposed to bankruptcy and failure. Thus we couple the threshold model to a two-strategy decision game in which agents decide to be protected (P) or not-protected (NP). The chosen strategy of each node $i$ at each time step $t$ is denoted by $d_i(t) \in \{ 0,1\}$, such that $d_i(t)=0$ if node $i$ is not-protected and $d_i(t)=1$ if it is protected. Therefore, in this model, nodes are in one of four states: active and not-protected (A,NP), active and protected (A,P), inactive and not-protected (I,NP), and inactive and protected (I,P) (see Fig.~\ref{model}(a)). We assume that the protected agents have higher threshold value than not-protected ones. Hence the threshold value of node $i$, denoted by $m_i(t)$, is updated according to the chosen strategy of that node at each step as follows,

\begin{equation}
	m_{i}(t)=\left\{ \begin{array}{l}
		{m \qquad \text{if $i$ is NP,}} \\
		{\sigma m \qquad \text{if $i$ is P,}}
	\end{array} \right.
	\label{mEqu}
\end{equation}
where $0<m<1$ and $\sigma$ is the threshold increase factor for protected nodes. In fact $\sigma$ shows the efficiency of the protection and determines how much more a protected node is resistant to failure than an unprotected one. We set $\sigma \ge 1$ with a possible maximum value of $\sigma_{max}=\frac{1}{m}$. Thus, $\sigma = 1$ implies that the protection is completely ineffective.

Moreover we assume that active but protected individuals are less likely to affect their neighbours rather than active and not-protected ones. For instance, in difficult economic conditions, insured companies suffer less damage due to the use of financial support, and as a result, they have a less destructive effect on the companies they are associated with. To this end, we use $\gamma \in (0,1)$ as a reduction factor in the effect that an active protected neighbor has on the activation of its neighbor node. Here we assume that the effect of active protected neighbors is half of the effect of active but not-protected ones and set $\gamma=0.5$ for simplicity.

Thus, the activation probability of node $i$ with degree $k_{i}$ at time $t$ is obtained as follow,

\begin{equation}
	\Gamma_{I \to A}^{i}(t) = H \Big{(} \frac{\sum_{j=1}^{N} \mathcal{A}_{ji} s_j(t)  (1-\gamma d_j(t))  }{k_i}  -  m_{i}(t)  \Big{)},
\end{equation}
in which $\mathcal{A}_{ji}$ are the elements of the adjacency matrix and $H(x)$ is the Heaviside function, where $H(x)=1$ if $x\geq 0$ and otherwise $H(x)=0$.
Figure~\ref{model}(b) represents a schematic of modified threshold model for the activation process of P and NP agents.

In the decision dynamics coupled to threshold model, nodes choose to become P or NP at each time $t$ respectively with probabilities $\Gamma_{NP\to P}$ and $\Gamma_{P\to NP}$, regardless of being active or not (Fig.~\ref{model}(c)). To this end, at each time step, the protection state of node $i$ is compared with a neighbour $j$ chosen randomly while we apply a synchronous update rule. If both have the same strategy, nothing happens. Otherwise, node $i$ imitates its neighbour strategy with probability $\Gamma_{P\to NP}$ ($\Gamma_{NP\to P}$) if it is initially protected (not-protected).
We consider these probabilities as the following Fermi functions:
\begin{eqnarray}
	\Gamma_{P\to NP}(t) = \frac{1}{1+e^{(\Pi_P(t)-\Pi_{NP}(t))}}, \\
	\Gamma_{NP\to P}(t) = \frac{1}{1+e^{(\Pi_{NP}(t)-\Pi_{P}(t))}},
\end{eqnarray}
where $\Pi_{P}$ and $\Pi_{NP}$ are the payoffs of being P and NP, respectively. We assume the payoff functions depend on the fraction of active nodes. The higher the fraction of active and protected (active and not-protected) nodes, the less profitable the choice of the protected (not-protected) strategy. Hence, the payoffs are defined as follows,
\begin{eqnarray}
	\Pi_{P}(t) = -c -\alpha \rho_{A,P}(t), \label{payoffP}\\
	\Pi_{NP}(t)= -\alpha \rho_{A,NP}(t).
	\label{payoffNP}
\end{eqnarray}
\begin{figure*}[t!]
	\includegraphics[scale=.55]{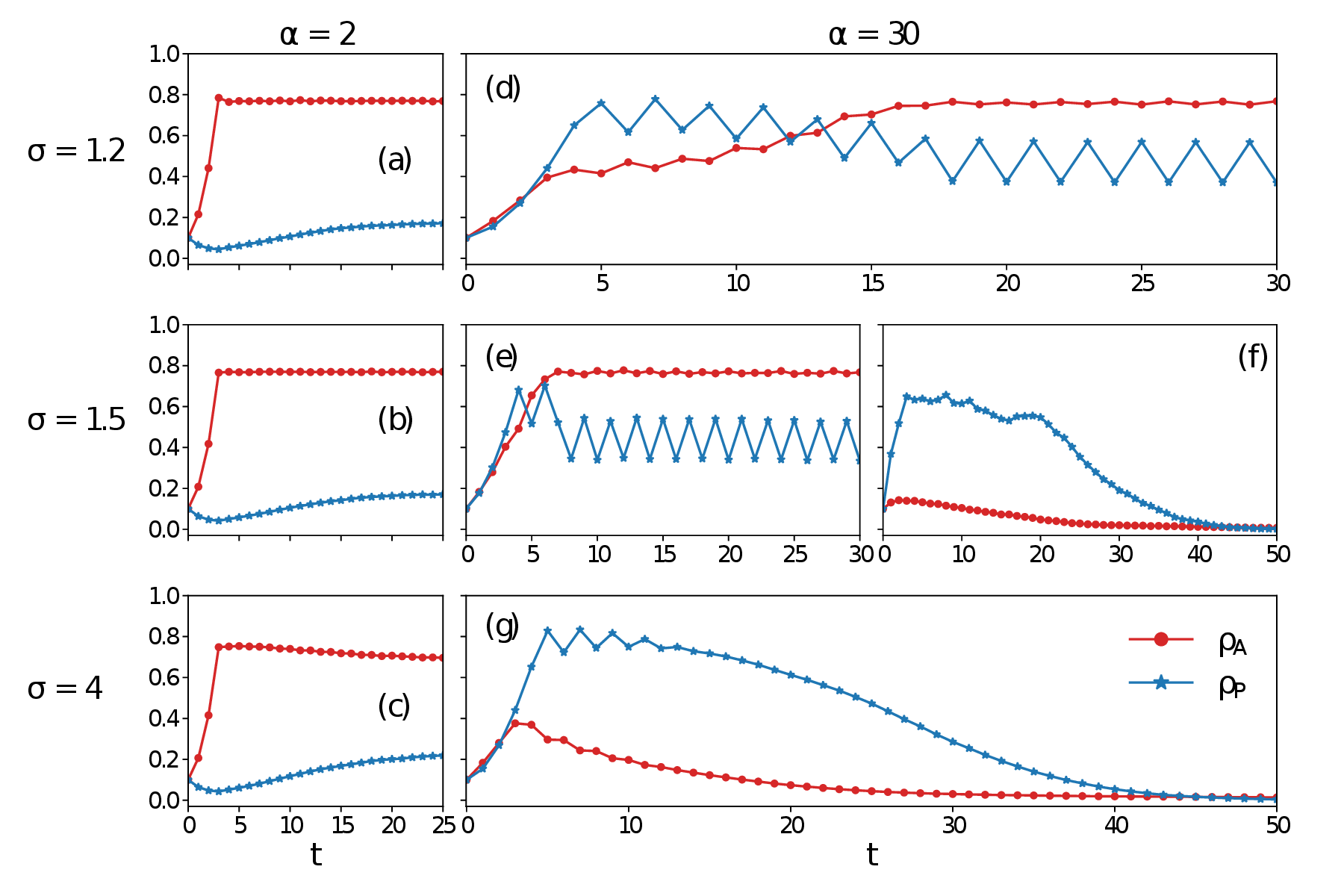}
	\centering
	\caption{Cascade size $\rho_A$ (red dots) and fraction of protected nodes $\rho_P$ (blue stars) versus time $t$ for two values of  $\alpha=2$ (left column) and $\alpha=30$ (right column). The threshold increase factor for protected nodes, $\sigma$, increases from top to bottom. The results are obtained on the ER network with $N=3000$, $\langle k \rangle=10$, $m=0.2$ and $\beta=0.3$.}
	\label{AS5}
\end{figure*}

According to Eqs.~(\ref{payoffP}) and (\ref{payoffNP}), nodes evaluate their strategic choice based on (i) the intrinsic cost that the individuals have to consider for choosing the protection (c), (ii) the cost of activation ‌(failure) which means how severe the consequences of being activated is ($\alpha$), and (iii) the perceived risk of facing a failure which depends on the fractions of active individuals who have adopted protected ($\rho_{A,P}$) or not-protected strategy($\rho_{A,NP}$). We can represent the fraction of (A,P) and (A, NP) agents in terms of the Delta function $\delta$, as the following equations:

\begin{eqnarray}
	\rho_{A,P}(t) &=& \frac{1}{N} \sum_{i=1}^{N} \delta (s_i(t) - 1) \delta ( d_i(t) -1), \label{rhoAPequ}\\
	\rho_{A,NP}(t) &=& \frac{1}{N} \sum_{i=1}^{N} \delta (s_i(t) - 1) \delta ( d_i(t)).
	\label{rhoANPequ}
\end{eqnarray}

Now we aim to see how the protective measures affect the size of the cascades.  To this end, we define the following variables of interest: the fraction of active nodes ($\rho_{A}$) and the fraction of protected ones ($\rho_{P}$) which are obtained as follows:

\begin{eqnarray}
	\rho_{A}(t)&=&\rho_{A,P}(t) + \rho_{A,NP}(t), \\
	\rho_{P}(t)&=&\rho_{A,P}(t) + \rho_{I,P}(t).
\end{eqnarray}
where, $\rho_{I,P}$ is the fraction of inactive and protected nodes, which similarly is represented by the Delta function as follows:
\begin{eqnarray}
	\rho_{I,P}(t) = \frac{1}{N} \sum_{i=1}^{N} \delta (s_i(t)) \delta (d_i(t) -1). \label{rhoIPequ}
	\label{rhoINPequ}
\end{eqnarray}
\begin{figure*}[t!]
	\includegraphics[scale=.55]{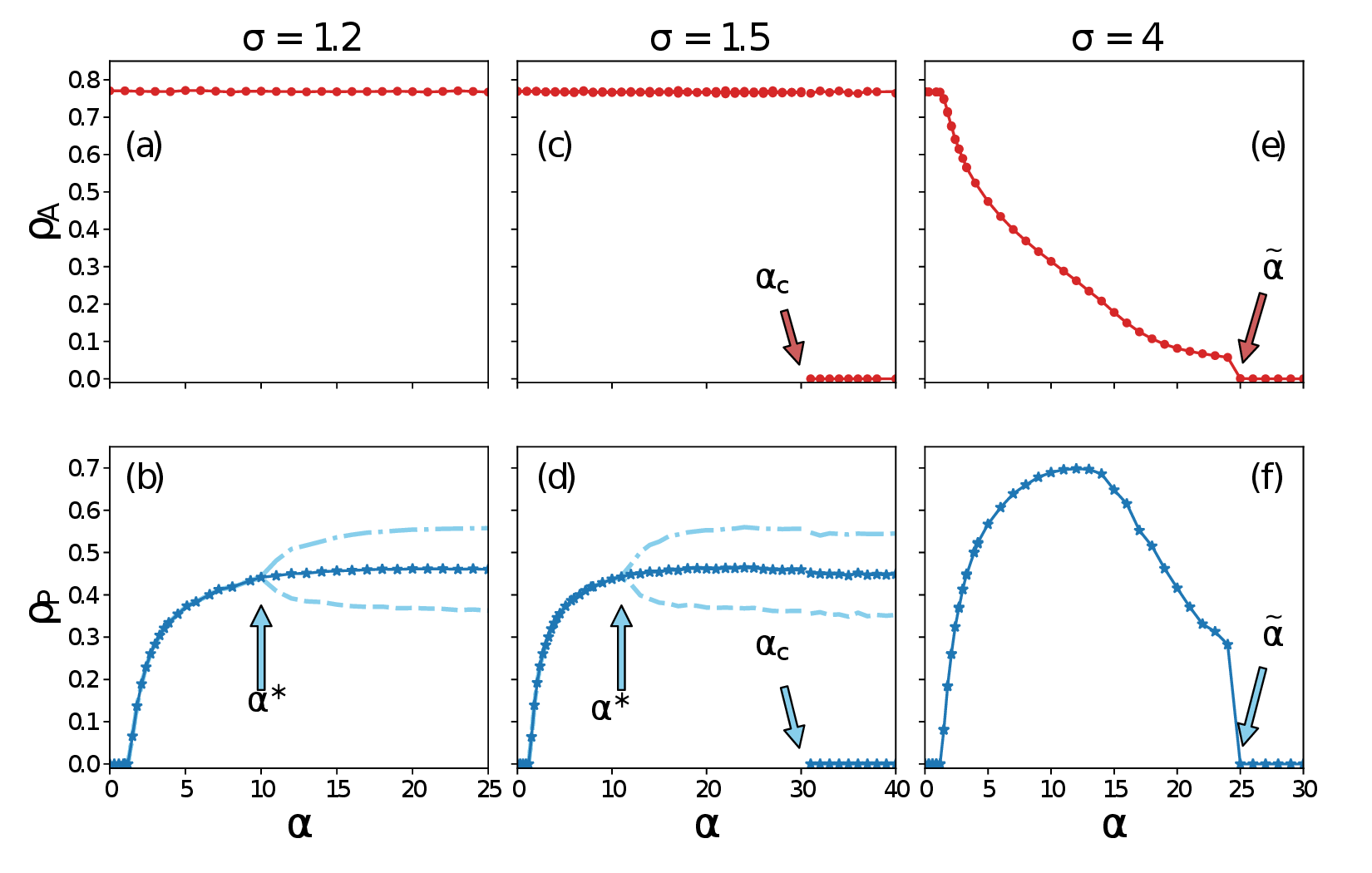}
	\centering
	\caption{Stationary values of the cascade size (red dots) and the fraction of protected nodes (blue stars) as a function of $\alpha$. The value of $\sigma$ increases from left to right. Bifurcation point $\alpha^*$, bistability point $\alpha_c$, and absorbing point $\overset{\sim}{\alpha}$ are shown in the figure. Dashed-dotted and dashed lines show the upper and lower turning points of stable cycles, respectively. Parameters are set as Fig.~\ref{AS5}. The results are averaged over 300 realizations }
	\label{AS4}
\end{figure*}
\subsection*{Simulation}
Once introducing dynamical rules of the model, now we analyse the behaviour arising from the interplay between two dynamics. To this end, we perform numerical simulations on Erd\H{o}s--R\'enyi (ER) random networks with $N=3000$ nodes and  mean degree $\langle k \rangle=10$.
The initial conditions are chosen as $\rho_{A}(0)=0.1$ and $\rho_{P}(0)=0.1$, such that 10 of nodes are selected to become active and the same fraction are set protected, randomly. To see how the decision-making and cascade dynamics coevolve, we study the behaviour of $\rho_{A}$ and $\rho_{P}$.
We choose the parameters in the range where the cascade is detected in the absence of protective dynamics
(i.e., when $\sigma=1$ and $\gamma=1$). Hence, we set $m=0.2$ and also consider $c=1$ in all simulations for the sake of simplicity.

\section*{Results} \label{sec3}
\subsection*{The role of the activation (failure) cost $\alpha$ and the threshold increase factor $\sigma$}
To see how to get into stationary state, we show the time evolution of cascade size $\rho_A$ and the fraction of protected nodes $\rho_P$ in Fig.~\ref{AS5}. The results are obtained for two values of $\alpha$ while we raise the threshold increase factor $\sigma$ from top to bottom.  According to Fig.~\ref{AS5}~(a)-(c), when the cost associated with activation ($\alpha$) is small, the prevention measures are useless such that a large cascade occurs, regardless of the value of $\sigma$. In this regime the system reaches a fixed point where the fraction of protected nodes is small.

Increasing $\alpha$, we observe different behaviours. As we can see in Fig.~\ref{AS5}~(d), for the small values of $\sigma$ (a low limit), the density of protected nodes increases and in the stationary state fluctuates periodically between two values. As a result of this behaviour, $\rho_{A}$ oscillates, but not to a large enough amplitude and gradually the system reaches an active phase where a relatively large cascade is obtained. On the other hand, for the large values of $\sigma$, the cascade collapses and consequently the density of protected nodes becomes zero (see Fig.~\ref{AS5}(g)). This could be thought as the behaviour of $\rho_{A}$ and $\rho_{P}$ in the high limit of $\sigma$, where the system reaches an absorbing phase. However, the most interesting behaviour occurs for the intermediate values of $\sigma$. In this regime, both the active phase, consisting a large cascade size (Figs.~\ref{AS5}~(e)) and the absorbing phase (Figs.~\ref{AS5}~(f)) are possible with the same probability, such that the bistability emerges.

Figure~\ref{AS4} shows the stationary values of $\rho_{A}$ and $\rho_{P}$ as a function of $\alpha$ for three different values of $\sigma$ (lower, intermediate, and upper limits), such that each regime undergoes a variety of phase transitions which we discuss below:

\begin{itemize}
\item $Lower~limit$: as we see in Fig.~\ref{AS4}~(a)-(b), increasing the failure cost $\alpha$ does not affect the cascade size and an active phase is observed. However, the protected fraction $\rho_P$ initially rises with $\alpha$. By further increasing $\alpha$, drives the system goes towards a supercritical Hopf bifurcation point $\alpha^*$, in which the fixed point for the protected fraction loses its stability while a 2-period cycle in which $\rho_{P}$ oscillates in time between two values emerges.

\item  $Intermediate~limit$: in this regime we observe a similar behavior with the lower limit for the small values of $\alpha$. However, for the values of $\alpha$ more than a critical value $\alpha_c$, the absorbing phase becomes stable too (see Fig.~\ref{AS4}~((c)-(d)). Indeed, a bistable region emerges, where the system reaches either the active phase or the absorbing phase.  Notably, $\rho_{P}$ oscillates between two values in the active phase, similar to its behavior for $\alpha^* < \alpha <\alpha_{c}$, while in the absorbing phase, both cascade size and protected fraction reach to zero.

\item $Upper~limit$: in this regime, increasing $\alpha$ leads to a rise in the density of protected nodes. This growth dramatically affects the size of the cascade and leads to a significant reduction in the fraction of active nodes. This drop continues until $\rho_A$ reaches zero at $\overset{\sim}{\alpha}$. The fraction of protected nodes also decreases after passing a maximum value and finally becomes zero at $\overset{\sim}{\alpha}$ (Fig.~\ref{AS4}~(e)-(f)).

\end{itemize}
\begin{figure*}[t!]
	\includegraphics[scale=0.53]{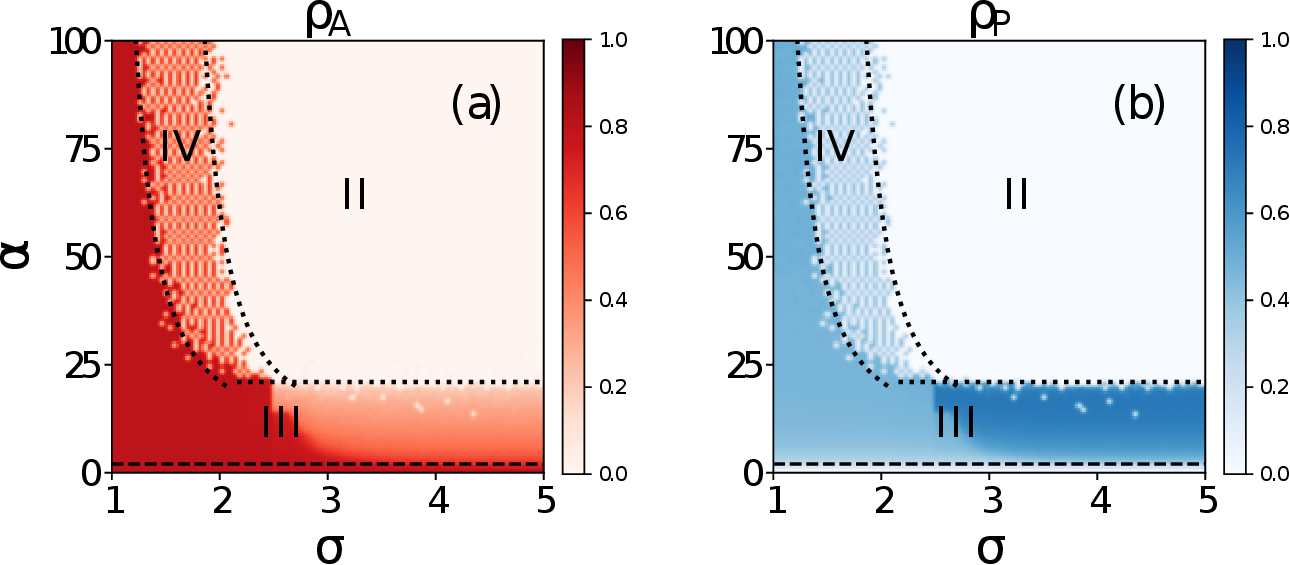}
	\centering
	\caption{Phase diagram for $(a)$ the cascade size and $(b)$ the fraction of protected nodes in the space $\alpha-\sigma$. Parameters are set as Fig.~\ref{AS5}. The checkered area distinguishes  bistable region. }
	\label{AS}
\end{figure*}
\begin{figure}[t!]
	\includegraphics[scale=0.5]{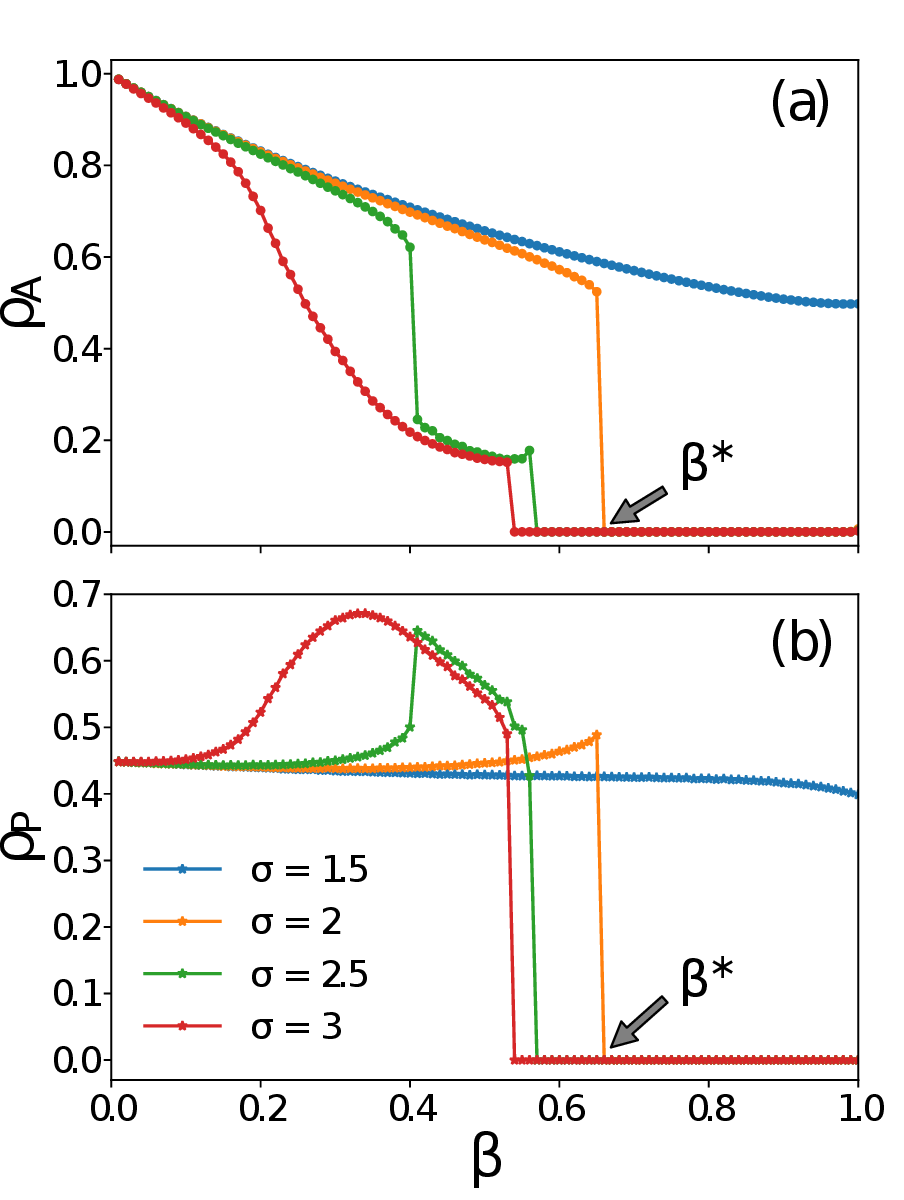}
	\centering
	\caption{Stationary values of $(a)$ the cascade size and $(b)$ the fraction of protected nodes as a function of $\beta$ for different values of protective efficiency $\sigma$ and $\alpha=10$. }
	\label{AS1}
\end{figure}
In Fig.~\ref{AS}, we show the phase diagram for the model in the space $\alpha-\sigma$. We can see different regimes, namely,

\begin{itemize}
\item  active state (I): wherein a large number of nodes are active while all the nodes are not protected. This happens, below the black dashed line.
\item inactive state (II): all nodes are inactive and also not protected ($\rho_{I,NP}=1$).
\item active and protected (III): a fraction of nodes are active and also some nodes are protected.
\item bistable (IV): where the active and inactive states exist at the same time.
\end{itemize}

\begin{figure*}[t!]
	\includegraphics[scale=0.53]{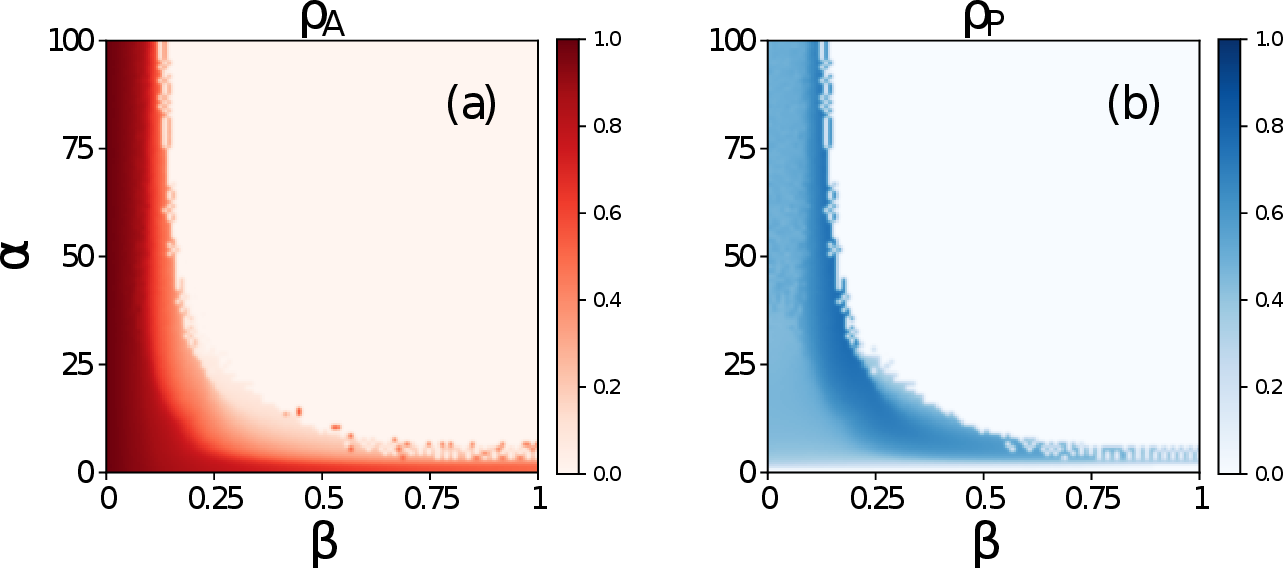}
	\centering
	\caption{Phase diagram for $(a)$ the cascade size and $(b)$ the fraction of protected nodes in the space $\alpha-\beta$ and for $\sigma=3$. }
\label{AS2}
\end{figure*}
\subsection*{The role of recovery probability $\beta$}

The active nodes return to inactive state with a constant probability $\beta$. Hence with increasing $\beta$, the cascade size decreases. However if the effectiveness of the ‌‌protection is large enough there is more desire to protect and consequently a further reduction in the cascade size is observed. We can see this behavior in Fig.~\ref{AS1}. With increasing the recovery probability $\beta$, the fraction of active nodes gradually decreases. However for $\sigma> 2$ the fraction of protected nodes grows and the density of active nodes finally jumps to zero in a transition point $\beta^*$. With the disappearance of activity, there is no reason for protection and the population of protected nodes drops also to zero. Hence, a transition occurs from the active and protected phase to the inactive phase at $\beta^*$ so that above this point the agents stop the protection and all are inactive. The value of the recovery transition point $\beta^*$ depends on the value $\sigma$. The more effective the protection (greater $\sigma$), the more the activation is reduced and $\beta^*$ is shifted towards lower values.

Figure \ref{AS1} shows phase transitions for $\alpha=10$. However, the transition point $\beta^*$ changes also with increasing $\alpha$. When the cost of activation $\alpha$ increases, there is a more requirement to protect and the fraction of protected agents increases, which causes less activation, and the fraction of active agents reaches zero at a lower transition point. In Fig.~\ref{AS2}, we show phase diagrams for the fraction of active agents $\rho_A$ and the fraction of protected agents $\rho_P$ as a function of $\beta$ and $\alpha$. As we can see, the cascade collapses at smaller values of the transition point $\beta^*$ with increasing $\alpha$. The result shows how much the agents' ability to rebuild is effective in preventing cascading failures. However with protection, the global cascade can be avoided even for low recovery probability values.

\section*{Discussion} \label{secCon}
Propagation of cascading failures can cause catastrophic events. Protective measures to prevent the agents from joining the cascade play an important role in controlling the cascade. In this work we studied the interplay between a decision game dynamic and the spreading of cascade. We considered the linear threshold model on the ER random networks for the dynamics of cascade and assumed that the nodes can be protected or not due to the perceived risk of failure and the costs associated with the protection measures. Using numerical simulations we demonstrated how the coevolutionary dynamics can significantly affect the size of cascade.

We assumed that the protection increases the threshold of the nodes such that the protected nodes are more robust against the failure. With increasing protective effects (increasing the threshold), the size of the cascade reduces further. In the case that the perceived risk of failures is sufficiently high, there is more desire for protection and consequently the size of protected nodes increases which has a significant effect on reducing the size of the cascade. Also we showed that with increasing the values of risk of failures, the sustained oscillations emerge in the fraction of protected nodes. In particular for the intermediate effectiveness of protective measures, the density of protected nodes and also the size of cascade show a bistable region in which both absorbing and active states are stable. Here we also considered the possibility of recovery for the active agents and found the transition point for the probability of recovery in which the size of cascade collapses.

In summary we proposed a simple model which shows the effective factors on cascade control. Paying insurance costs by individuals or companies can be effective in some situations and prevents irreparable damages. Also the ability to recover and return to normal conditions can prevent the occurrence of major failures. In particular, the effect of recovery is enhanced by adding protection. Other factors such as heterogeneity of thresholds, network structure and correlation between components as well as local information in the payoff functions can affect the results which are a challenge for future work.








\end{document}